# A key role for unusual spin dynamics in ferropnictides

I.I. Mazin[1] & M.D. Johannes[1]

[1]*Center for Computational Materials Science, Naval Research Laboratory Laboratory, Washington, D.C. 20375, USA*

**The 2008 discovery of superconducting ferropnictides with $T_c$~26K-56K introduced a new family of materials into the category of high $T_c$ superconductors. The ongoing project of understanding the superconducting mechanism and pairing symmetry has already revealed a complicated and often contradictory underlying picture of the structural and magnetic properties. There is an almost unprecedented sensitivity of the calculated magnetism and Fermi surface to structural details that prohibits correspondence with experiment. Furthermore, experimental probes of the order parameter symmetry are in surprisingly strong disagreement, even considering the relative immaturity of the field. Here we outline all of the various and seemingly contradictory evidences, both theoretical and experimental, and show that they can be rectified if the system is assumed to be highly magnetic with a spin density wave that is well-defined but with magnetic twin and anti-phase boundaries that are dynamic on the time-scale of experiments. Under this assumption, we find that our calculations can accurately reproduce even very fine details of the structure, and a natural explanation for the temperature separation of structural and magnetic transitions is provided. Thus, our theory restores agreement between experiment and theory in crucial areas, making further cooperative progress possible on both fronts. We believe that fluctuating magnetic domains will be an essential component of unravelling the interplay between magnetic interactions and superconductivity in these newest high $T_c$ superconductors.**

Two decades of studying the high $T_c$ cuprates have not produced a full understanding of their properties, but in two major aspects there has been impressive progress: first, the symmetry of the superconducting order parameter is known: it is one-band $d_{x^2-y^2}$. Second, there is a reasonably good understanding of the parent (undoped) compounds: these are strongly correlated Mott-Hubbard insulators with local magnetic moments at the Cu sites that interact locally through the superexchange mechanism. The essential physics happens on the local level and is largely captured by the dynamic mean field theory which is local by construction.

Neither of these things is known regarding the newly discovered ferropnictides [1]. Perhaps more importantly, it is known that the basic facts outlined above for cuprates do *not* apply to pnictides. The parent compounds are metallic, rather than insulating, and in some cases exhibit weak magnetism rather than robust, localized moments. When it manifests, this magnetism is strongly affected by minor changes in the crystal structure and, in particular, by changes in the electrically inert rare earth separator layer. This is in sharp contrast to the cuprates where magnetism is completely insensitive to iso-electronic rare earth substitutions. Andreev reflection experiments, penetration depth measurements, and photoemission all exclude the possibility of gap nodes, thereby eliminating the standard *d*-wave superconductivity as we know it in cuprates.

One of the most mysterious differences between the cuprates and ferropnictides is the way in which standard density functional theory (DFT) band structure calculations fail to describe them. In cuprates, DFT calculations do not well describe the local Coulomb correlations that enhance the tendency toward local moment formation and consequently, barely magnetic or fully non-magnetic solutions result. Conversely, in the ferropnictides, DFT calculations invariably converge to a spin density wave (SDW) state with magnetic moments significantly *larger* than experiment (1.5-2 $\mu_B$), whether in doped or undoped materials. Experimentally, antiferromagnetism is observed only at very low doping levels and is often very weak. The overestimation of magnetic strength compared to experiment is rare in DFT and definitively removes the ferropnictide family from the strongly correlated regime of the cuprates. Magnetic moments in both calculation and experiment appear to be very soft and change dramatically as a function of seemingly minor details. Among the many unexpected features found in calculational results, two are particularly striking. First, enforcing collinear alignment of all spins (a ferromagnetic ordering) destroys magnetism nearly entirely. To induce a FM moment comparable to the one calculated in the AFM ground state, an external field of ~2 kT must be applied. Second, the total energies so obtained cannot be fit to a Heisenberg model with first, second or third neighbour interactions.[2] This undoubtedly indicates that magnetism in this compound is itinerant and requires coherence on the order of at least several lattice parameters in order for a magnetic state to form.

We propose that these failures are due to an underlying ground state that is strongly magnetic, but with fluctuating domain boundaries that preclude its experimental detection. This rather simple assumption not only brings computational and experimental results into startlingly good agreement, but also provides a natural explanation for many experimental observations that otherwise appear incongruent. Below we summarize the current state of affairs with respect to what is known about the ferropnictides, first experimentally and then theoretically, pointing out where contradictions arise. We then show how a consistent picture can be formed by considering various features of our postulated magnetic state.

**Experiment:** (1) Undoped LaFeAsO has probably been studied in more depth than any other ferropnictide. It is well established[3] that it experiences a very weak structural distortion at ~150 K, followed by formation of a SDW at ~140 K whose amplitude grows with cooling up to 0.3-0.4 $\mu_B$. The magnetic moments form stripes: nearest neighbor spins are aligned along one direction and anti-aligned along the other, and stripes in adjacent planes are anti-aligned (See Fig. 2a,b). Such an ordering might, in principle, be explained by competition between AFM nearest- and next-nearest-neighbor couplings (if the former were about twice the latter). The surprisingly small observed ordered moment would then be attributed to frustration. However, this explanation cannot be right: magnetic frustration would lead to large static local moments near any crystallographic defect or impurity (*cf.*, *e.g.*, $LiV_2O_4$) that would be detectable by μSR or Mössbauer experiments. Yet, both of these experiments find magnetic moments similar to, or even smaller than, the ones measured with neutrons[3,4,5], even in bad and off-stoichiometric samples. In addition, as discussed below, such an explanation flatly contradicts first principles calculations.

(2) Iso-electronic manipulations within the (electrically charged, electronically inert) LaO layer affect the magnetism in an unexpectedly strong manner. Substituting La by Nd suppresses the Fe moment to a barely detectable level until the Nd spins finally order at 2K[6]. Most surprisingly, the SDW that forms in the Fe layer at that temperature has magnitude of 0.9 $\mu_B$, two to three times larger than in LaFeAsO. A Ce substituted compound, like La, orders at about 140K - but with a moment magnitude of 0.6 $\mu_B$.[7] The Neel temperature of Ce or Nd subsystems remains very low, on the order of 1-2 K, indicating an absence of noticeable magnetic coupling between the rare earth and Fe moments. In fact, AFM ordering of Ce has been seen in superconducting samples[8], reinforcing that Ce f electrons are not coupled to Fe d states. This contrasts with the $YBa_2Cu_3O_4$ family in which, except for Pr, no rare earths couple with the Cu d-electrons, their ordering temperature remains low and they do not affect superconductivity. On the other hand, Pr does couple with the metallic states, orders at temperature that is an order of magnitude higher, and destroys superconductivity. It is worth noting that each rare earth ion projects onto the center of square Fe plaquettes with equal numbers of up and down spins, so that within the Heisenberg exchange model ,they do not couple with Fe moments at all. Finally, in $BaFe_2As_2$, which has no magnetic species besides Fe, the neutron measured moment is also close to 0.9 $\mu_B$ [9] (although Mössbauer spectra suggest a twice smaller moment[4,10]).

(3) Contrary to initial expectation, the resistivity does not increase, or increases very little upon the onset of the SDW (which presumably gaps most of the Fermi surface), and then drops precipitously with further temperature lowering. The in-plane to out-of-plane transport anisotropy does not change at all for the entire temperature range. This can only be interpreted as the rapid removal of some (isotropic) scattering channel at the Neel temperature that affects the overall carrier density only slightly.[11] A sharp drop[3] of the Seebeck coefficient right below the transition is even stronger evidence against a sharp drop in the carrier concentration, but is quite consistent with a rapid change of the electron and hole relaxation time ratio.

(4) The Seebeck coefficient for the doped compounds is anomalously large[12,13], in excess of 100 $\mu$V/K, with a well-expressed maximum at low temperatures (~100 K). This is typical of doped semiconductors rather than of sizeable Fermi surface metals, as these compounds are usually assumed to be.

(6) Experiments that directly or indirectly probe the superconducting gap and its symmetry have produced a variety of results. Two types of experiment, ARPES[14,15] and PCAR[16], indicate the presence of several nodeless gaps. The temperature behavior of some gaps is reminiscent of the *s*-wave BCS superconducting kind, but others show an odd *T*-dependence that suggests a different (possibly magnetic?) order parameter. Multiple penetration depth experiments also suggest nodeless gaps, and Scanning SQUID Microscopy[17] excludes *d*-wave pairing, but NMR relaxation rate measurements show a power law behavior all the way down to 0.1 $T_c$, that suggests low energy relaxation by something besides than one-electron excitations[18,19]. Interestingly, superconductivity arises in many members of the ferropnictide family as a function of doping, but the doping level itself is actually irrelevant to the onset of pairing. The true controlling factor is the suppression of magnetism, which may be caused by doping, but can also be brought about by alternate means. It has been shown that when the long range order is suppressed by pressure, superconductivity appears, regardless of whether or not the compound is doped [20,21].

*Theory:* (1) First principles calculations predict a doping-independent metallic ground state with a large amplitude SDW ($M$=1-2 $\mu_B$/Fe) with the same ordering pattern as observed in experiment[2,22]. As opposed to most antiferromagnets, from Cr to NiO, the results of such calculations cannot be presented in terms of local moments interacting via pairwise exchange interaction. And unlike typical antiferromagnets, ferropnictides cannot be forced into a metastable ferromagnetic state, though in the GGA formalism, for a subset of possible structural parameters, a FM state with a very small moment (never comparable to the AFM solution) can be realized. If stabilization of the FM state is forced with an external field in order to compare it with AFM states, it appears that the corresponding energy differences cannot be mapped onto a two-nearest-neighbor Heisenberg Hamiltonian. Finally, exchange parameters calculated as the second derivative of the total energy with respect to the spin misalignment angle strongly depend (even in sign!) on the underlying ordering pattern. For instance, for the actual antiferromagnetic stripe ordering, the calculated exchange constant between anti-aligned neighbors is 550 K and for aligned neighbors is -80 K. Obviously, for a checkerboard ordering (all nearest neighbors anti-aligned), both constants must be equal. Finally, the calculated nonmagnetic state is stable against small perturbations,[22] that is, the spin susceptibility does not diverge at the wave vector required for the observed SDW; nonetheless, a finite amplitude SDW is substantially more stable than the nonmagnetic state. The origin of the magnetic stabilization energy can be traced down to one-electron energies. AFM (but not FM) ordering opens a pseudogap at the Fermi level, substantially lowering the band energy. As seen in Fig. 1, along one of the two crystallographic directions, the electron pockets disappear, and along the other, shrink drastically. The two hole pockets split, with one of them closing and the other strongly diminished.

(2) Structural optimization without accounting for magnetism leads to Fe-As or Fe-P bond lengths that are much shorter than those observed experimentally (by up to 0.15 Å). On the other hand, allowing for full spin polarization leads to pnictogene positions that are very close to the experiment (errors less than 0.03 Å), but yield a very large magnetic moment ( ~1-2 $\mu_B$/Fe) not seen in experiment. This holds for all three major modifications of the ferropnictide family: LaFeAsO, BaFe$_2$As$_2$, and LaFePO. The last type (LaFePO) is fully non-magnetic experimentally and the calculated ground state moment is much smaller (0.6 $\mu_B$) in comparison to the other two types. Correspondingly, the error in the calculated Fe-P bond length calculated in the nonmagnetic (NM) case is also much smaller (0.05 Å) than in the two other families[23] (Table 1). It appears that the discrepancy between the calculated Fe-pnictogen bond length and the experiment is directly proportional to the calculated ground state magnetic moment! Furthermore, a calculation in which the Fe ion carries a reasonably large moment leads to a correctly reproduced pnictogen position, regardless of the particular ordering pattern that is established. We have verified that optimizing the As and La positions within the checkerboard AFM structure, which is entirely different from the observed SDW, results in practically the exact same coordinates as using the actual SDW ground state magnetic pattern.

(3) The structural distortion observed in experiment cannot be reproduced using non-magnetic calculations. Establishing the AFM stripe phase again resolves the problem and, as observed in experiment[24,25] and previously obtained computationally[25,26], we obtain the relative contraction of the Fe-Fe distance between parallel spin neighbors compared to anti-parallel neighbors correctly using a full structural relaxation as described in Ref. 2. In contrast to the Fe-As distance, however, obtaining the structural distortion requires that the *correct* SDW pattern be applied. In the AFM checkboard pattern, just as in nonmagnetic calculations, Fe-Fe neighbours along both directions are equidistant and the ground state structure remains undistorted. This suggests that the structural transition is in some way driven by the AFM stripe magnetism, a fact which does not square with the observed occurrence of the magnetic transition at a lower temperature than the structural one. It is worth noting that while it is not necessarily the strongest superexchange interaction, the one due to direct Fe-Fe overlap is most sensitive to the Fe-Fe distance. In other words, if the magnetic ordering were being driven by superexchange, the antiferromagnetic bonds would contract, in contrast to the observed and calculated expansion.

(4) The calculated anisotropy in the squared plasma frequency, which corresponds to the resistivity anisotropy in the isotropic-scattering approximation, is about five times larger in the stripe AFM phase than in the nonmagnetic phase, in contradiction with the experimental observation that the onset of the SDW does not change the anisotropy. The calculated *value* of the squared plasma frequency (that is, effective number of carriers) in the stripe AFM phase is one order of magnitude *smaller* than in the nonmagnetic phase, while in the experiment the resistivity of the high-temperature phase extrapolates at $T=0$ to a number at least twice *larger* than the actual low-$T$ resistivity in the AFM phase[15].

(5) In the energy-independent relaxation time approximation, the calculated Seebeck coefficient in the nonmagnetic phase is just a few $\mu$V/K, and has the wrong sign compared to experiment. Yet again, allowing for full spin polarization (1.8 $\mu_B$/Fe) in the AFM phase brings it into reasonable range[23] of the experimental value of ~ -100 $\mu$V/K.

*A dynamic spin density wave system*. It is possible to reconcile almost everything known about the ferropnictide family of compounds and their properties, as laid out above, by assuming that the underlying system truly is magnetic. First, we assume that the actual ground state of an ideal system at $T$=0 is AFM stripe SDW with a large magnitude that is close to the DFT-calculated one, but which is likely to be reduced to ~1 $\mu_B$ by conventional zero-point spin fluctuations. The magnetic energy associated with this magnetic moment is responsible for expanding the Fe-As bond and driving the orthorhombic distortion, and thus computational and experimental structures are in excellent agreement. However, given the relatively small energy difference between the AFM stripe magnetic structure and other AFM patterns (cf. Refs.2,27), a large number of antiphase boundaries will form, even at very low temperatures (see Fig. 2a). Moreover, since the interlayer magnetic interactions are extremely small (our calculations put an upper bound of a few K for the interlayer exchange energy), the concentration of stacking faults along the $z$ direction should be exceedingly large. Without a more detailed theory and more specific information, it is difficult to quantify the dynamics of these defects (antiphase boundaries and stacking faults), but since there is no clear mechanism for pinning, they probably do not fully freeze in even at relatively low temperatures. The interlayer magnetic coherence in particular is likely very fragile, and, correspondingly, a true long range order detectable by neutrons would occur only in a small fraction of the sample (or in none at all) and would be subject to suppression by doping. Furthermore, fluctuations that correspond to the SDW wave vector along the two directions, *i.e*. $Q_x =(\pi,0)$ and $Q_y=(0,\pi)$, will suppress long range order in two dimensions, but will be strongly reduced by any three-dimensional interaction. Such fluctuations likely play a large factor in the unusual sensitivity of the magnetic transtion to the rare earth layer that effectively controls the three-dimensionality of the compound. In other words, most of the of Fe ions will be part of an SDW domain at any given moment of time, but will flip their spin every time a domain wall passes through that site. On the time scale of µSR or Mössbauer spectroscopy ($10^{-8}$ sec or slower) these sites will be observed to have substantially reduced moments or appear nonmagnetic altogether.

According to this scenario, $T_N$ (where the SDW order becomes detectable) can be understood as the temperature below which the antiphase boundaries are pinned by the establishment of three-dimensional coherency. For $T_N<T<T_s$ (where $T_s$ is the structural transition temperature), there is no long-range magnetic order due to the now dynamic antiphase boundaries, but the magnetic x/y symmetry is violated, because each magnetic domain has the same orientation, despite numerous misalignments of domains (See Fig. 2a). This symmetry breaking obviously induces the crystallographic symmetry lowering that is observed in experiment and calculations. At $T_s$, the system moves from a state dominated by antiphase boundaries where little twinning exists to a state in which the main defects are twin domain walls (Fig. 2b). According to recent data[3], twinning is incomplete all the way up to $T$~200 K, with an imbalance between x- and y-oriented AFM stripe domains remaining at all lower temperatures. At higher temperatures, the concentration of the two domain orientations is the same, and the global symmetry is tetragonal. Twinning rapidly disappears upon cooling below $T_s$ and is nearly (though still incompletely, according to Ref. 3) absent below $T_N$, at which temperature the 3D coherency first sets in. Since the twin and antiphase boundaries are electronically different, they scatter electrons differently so that both transitions are expected to be observable in transport properties. Indeed, the differential resistivity, $d\rho(T)/dT$, shows a sharp change of slope[5] at $T_s$ and a peak at $T_N$. The well-documented rapid drop in resistivity below $T_N$ in single crystals is thus associated with freezing of the antiphase domain walls. Note that in this model, the carrier concentration does not change drastically (neither does the band structure) at either $T_s$ or $T_N$. Rather, it is the relaxation rate that changes. This explains the surprising invariance of the resistivity anisotropy over the entire temperature range, including the onset of the SDW. It is worth noting that the domain picture as visualized by Fig. 2(a,b) is a useful, but possibly oversimplified picture. An alternative way to view the situation is that above $T_s$, SDW fluctuations with q=$Q_x$ and q=$Q_y$ have the same weight, while below $T_s$, fluctuations with one particular wave vector dominate, thus breaking the x/y symmetry.

Since the carrier scattering defects are magnetic in origin, interesting magnetoelastic effects can be expected in this system. A large magnetoresistance has in fact been observed in $BaFe_2As_2$ [11] at $T \lesssim 100$ K. The small carrier concentration in the magnetic phase also helps to explain the large thermopower. Finally, the mysterious sensitivity of the magnetic ordering (but almost no other properties) to the character of the inert space-filling layer (LaO, CeO, SmO, Ba, Eu etc.) finds a natural explanation: the establishment of long-range magnetic order is a 3D phenomenon. While most of the physical properties of the system are defined by the formation of SDW domains in individual FeAs planes, a detectable long-range ordering and a transition from the slow dynamics of domain walls (zero net magnetization on any given site over a long period of time) to the freezing of the domain walls requires 3D coherency. This last process is naturally sensitive to the properties of the filling layer and, in particular to the presence or absence of magnetic moments there, despite the near complete lack of magnetic interaction between the rare earth and Fe ions.

It is also tempting to associate some of the gaps observed in PCAR[16] and ARPES[14,15] with a dynamic SDW (pseudo)gap. In fact, the authors of several experimental reports already favor such an interpretation. At the present stage, the dynamic magnetic domain scenario remains a hypothesis, albeit an attractive one that unites theory, experiment and previously irreconcilable observations. The goal of this paper is to attract attention of experimentalists and theorists to this possibility. Currently, no other model consistently explains the entire body of experimental and computational evidence. It remains to be seen how unusual, topological excitations such as antiphase and twin domain boundaries may affect and/or possibly cause, the high temperature superconductivity in ferropnictides. It is worth recalling that very different, but also topological excitations have been intensively discussed in cuprate superconductors[28].

[1] Kamihara, Y., Watanabe, T., Hirano, M., and Hosono, H. Iron-Based Layered Superconductor La[$O_{1-x}F_x$]FeAs ($x$ = 0.05-0.12) with $T_c$ = 26 K. *J. Am. Chem. Soc.* **130**, 3296-3297 (2008).

[2] Mazin, I.I., Johannes, M.D., Boeri, L., Koepernik, K., and Singh, D.J. The challenge of unravelling magnetic properties in ferropnictides. Phys. Rev. B (in press).

[3] McGuire, M.A. *et al*. Phase transitions in LaFeAsO: structural, magnetic, elastic, and transport properties, heat capacity and Mössbauer spectra. arXiv:0806.3878

[4] Rotter, M. *et al*. Superconductivity at 38 K in the iron arsenide ($Ba_{1-x}K_x$)$Fe_2As_2$. Phys. Rev. Lett. (submitted).

[5] Klauss, H. –H. *et al*. Commensurate Spin Density Wave in LaOFeAs: A Local Probe Study. arXiv:0805.0264

[6] Chen, Y. *et al*. Magnetic Order of the Iron Spins in NdOFeAs. arXiv:0807.0662.

[7] Zhao, J. *et al*.. Structural and magnetic phase diagram of CeFeAs$O_{1-x}F_x$ and its relationship to high-temperature superconductivity arXiv:0806.2528

[8] Chen, G.F. *et al*. Superconductivity at 41K and Its Competition with Spin-Density-Wave Instability in Layered Ce$O_{1-x}F_x$Fe As. Phys. Rev. Lett. **100**, 247002 (2008)

[9] Huang, Q. *et al*. Magnetic order in BaFe$_2$As$_2$, the parent compound of the FeAs based superconductors in a new structural family arXiv:0806.2776

[10] Aczel, A.A , Muon spin relaxation studies of magnetic order and superfluid density in antiferromagnetic NdOFeAs, BaFe$_2$As$_2$ and superconducting (Ba,K)Fe$_2$As$_2$, arXiv:0807.1044

[11] Wang, X.F., *et al*. Growth and Anisotropy in transport properties and susceptibility of single crystals BaFe$_2$As$_2$. arXiv:0806.2452

[12] Sefat, A. S. *et al*. Electronic correlations in the superconductor LaFeAsO$_{0.89}$ F$_{0.11}$ with low carrier density Phys. Rev. B **77**, 174503 (2008)

[13] Pinsard-Gaudart, L., Berardan, D., Bobroff, J., and Dragoe, N. Large Seebeck coefficients in Iron-oxypnictides: a new route towards n-type thermoelectric materials. arXiv:0806.2751

[14] Zhao, L. *et al*. Unusual Superconducting Gap in (Ba,K)Fe$_2$As$_2$ Superconductor. arXiv:0807.0398

[15] H. Ding *et al*. Observation of Fermi-surface-dependent nodeless superconducting gaps in Ba$_{0.6}$K$_{0.4}$Fe$_2$As$_2$, EuroPhys. Lett. **83**, 47001 (2008)

[16] Gonnelli, R.S. *et al*. Coexistence of superconducting and magnetic energy scales in the superconducting state of LaFeAsO$_{1-x}$F$_x$. Nature (submitted)

[17] Hicks, C.W. *et al*. Limits on the Superconducting Order Parameter in NdFeAsO$_{1-x}$F from Scanning SQUID Microscopy. arXiv: 0807.0467

[18] Mukuda, H. *et al*. $^{75}$As NQR/NMR Studies on Oxygen-deficient Iron-based Oxypnictide Superconductors LaFeAsO$_{1-y}$ ( y= 0, 0.25, 0.4) and NdFeAsO$_{0.6}$ arXiv:0806.3238

[19] Nakai, Y., Ishida, K., Kamihara, Y., Hirano, M. and Hosono, H. Evolution from Itinerant Antiferromagnet to Unconventional Superconductor with Fluorine Doping in

La($O_{1-x}F_x$)FeAs Revealed by $^{75}$As and $^{139}$La Nuclear Magnetic Resonance. J. Phys. Soc. Japan. **77** (in press).

[20] Park, T. et al. Pressure-induced superconductivity in single crystal $CaFe_2As_2$. arXiv: 0807.0800

[21] Alireza, P., et al. Superconductivity up to 29 K in SrFe2As2 and BaFe2As2 at high pressures. arXiv: 0807.1896

[22] Yin, Z.P. et al. Electron-Hole Symmetry and Magnetic Coupling in Antiferromagnetic LaOFeAs. Phys. Rev. Lett. (in press)

[23] These results have been obtained using the full-potential LAPW method, in the WIEN2k implementation, as described in Ref. 2.

[24] Zhao, J. et al. Spin and Lattice Structure of Single Crystal $SrFe_2As_2$. arXiv:0807.1077

[25] Jesche, A. et al. Strong coupling between magnetic and structural order parameters in $SrFe_2As_2$. arXiv:0807.0632

[26] Yildirim, T. Origin of the ~150 K Anomaly in LaOFeAs; Competing Antiferromagnetic Superexchange Interactions, Frustration, and Structural Phase Transition. arXiv: 0804.2252

[27] Ishibashi, S., Terakura, K., and Hosono, H. A Possible Ground State and Its Electronic Structure of a Mother Material (LaOFeAs) of New Superconductors. J. Phys. Soc. Japan. **77**, 053709 (2008).

[28] Zaanen, J., and Gunnarsson, O. Charged magnetic domain lines and the magnetism of high-*Tc* oxides. Phys. Rev. B **40**, 7391(1989)

Table I: Calculated and experimental As positions and Fe moments

|  | Exp | GGA-NM | | GGA-AF | | GGA |
| --- | --- | --- | --- | --- | --- | --- |
|  | $z_{As}$ | $z_{As}$ | error | $z_{As}$ | error | (calc) |
| LaFeAsO | 0.65133 | 0.6375 | 0.12 Å | 0.6478 | 0.03 Å | 2.06 $\mu_B$/Fe |
| BaFe$_2$As$_2$ | 0.3545 | 0.3448 | 0.13 Å | 0.3520 | 0.03 Å | 1.97 $\mu_B$/Fe |
| LaFePO | 0.6339 | 0.6225 | 0.05 Å | 0.6254 | 0.03 Å | 0.60 $\mu_B$/Fe |

Figure 1: **The Fermi surface of LaFeAsO**. (a) Nonmagnetic Fermi surface. (b) The same, but with a rigid exchange splitting between electron and holes (~ ± 0.01 Ry). (c) The same as (b), but folded down so as to match the SDW Brillouin zone. (d) The Fermi surface in the calculated SDW ground state.

Figure 2: **Representative ferropnictide in-plane magnetic domains**. (left) $T<T_N$ anti-phase domains only, (right) $T_N<T<T_s$ anti-phase and twin domains. The green lines show antiphase domain walls and the brown lines the twin boundaries.

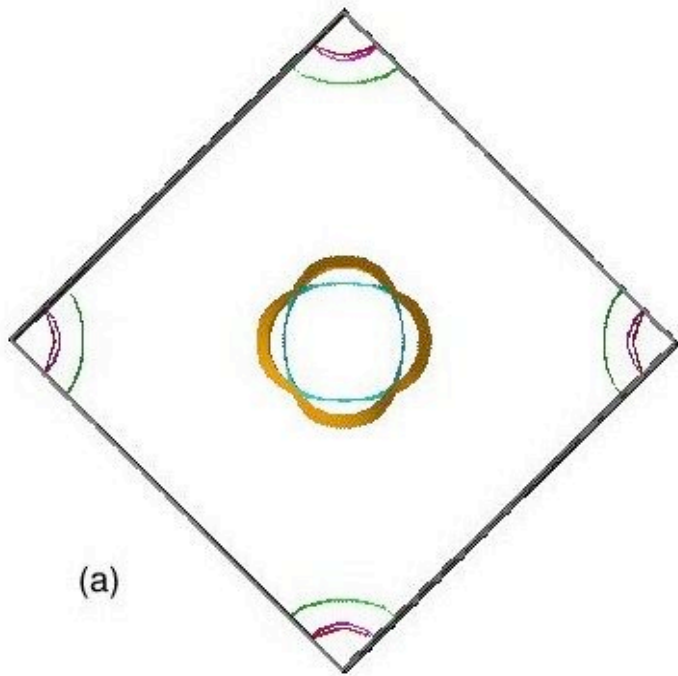
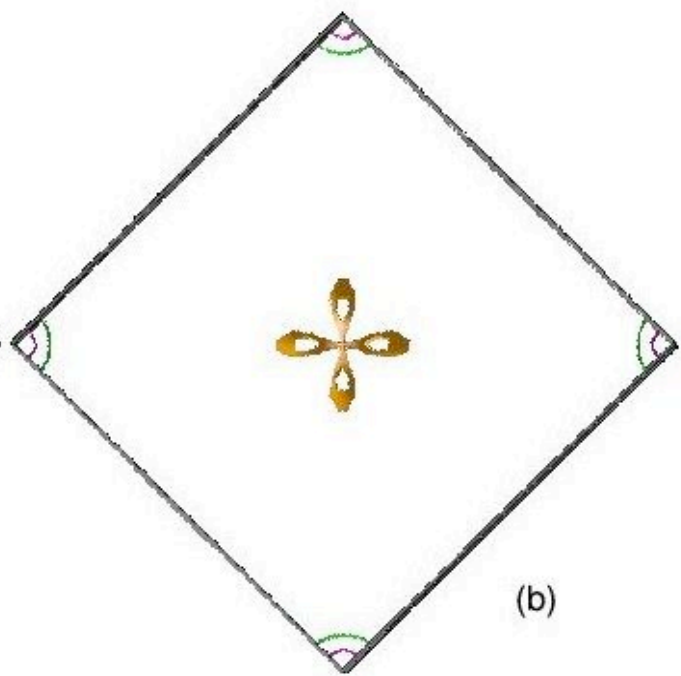
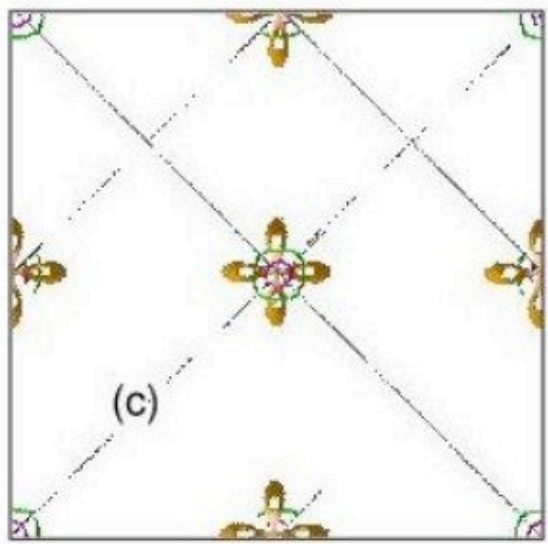
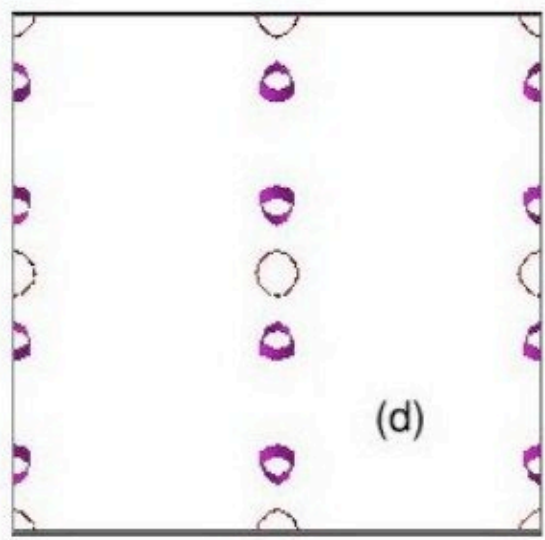